\documentclass[preprint,12pt,english,brazil]{elsarticle}

\usepackage{amssymb}

\setcitestyle{authoryear,open={(},close={)}}  
\journal{Communications in Nonlinear Science and Numerical Simulation}

\usepackage{multirow}

\begin{document}

\begin{frontmatter}


\title{Forecast modeling a time series of water reservoir levels using exponential smoothing method}
 \author{Souza, Lydiane F.\corref{cor1}\fnref{label1}}
\cortext[cor1]{\ead{lydianede1@hotmail.com }}
\affiliation{organization={Center for Exact Sciences and Technologies,
Federal University of Western Bahia}, addressline={Rua da Prainha, 1326, Morada Nobre}, city={Barreiras},postcode={47810-047},state={Bahia},country={Brazil}}

\title{}


\author{}


\begin{abstract}
Exponential smoothing is a time series forecasting method that presents the forecast based on trend and seasonality components. In this work, we study the behavior of two time series that describe the level of the water reservoirs of the Descoberto and Santa Maria dams. We trained the fifteen models present in the Pregels taxonomy, the criterion for choosing the model consists of the model with the lowest Akaike information criterion. The results indicate that the exponential smoothing model with damped additive trend and additive seasonality best describes both time series.
\end{abstract}




\end{frontmatter}


\section{Introduction}
\label{}
Exponential smoothing (ES) methods are widely used in demand forecasting \citep{Gardner}, production and inventory in the business area \citep{Sigerar, Ferbartrata}, telecommunications data \citep{GardnerJr} and is also used in weather forecast \citep{sensors,appli}. These forecasting methods have been widely used since 1950 because their mathematical formulation is simple, requires little computational time, and has results with reasonable accuracy. The ES methods have 15 variations, depending on the trend and seasonality used in the forecast; the best known are simple exponential smoothing \citep{Brown}, Holt method \citep{Holt} and Holt-Winters method \citep{Winters}.

This work will use the ES methods to obtain the best prediction model of the historical series of water levels of the two main water reservoirs that supply the Federal District in Brazil. The reservoirs of the Descoberto and Santa Maria dams are the oldest water reservoirs in the Federal District. The historical series of the level of the Descoberto and Santa Maria reservoirs used in this research starts on April 1987 and ends on October 2021\citep{dataset}; the historical series of the Descoberto dam has 12307 data, and the Santa Maria dam has 12399 data.

The study's objective is to obtain the most suitable model for each reservoir and predict this series's behavior for the coming years to encourage public policies for the preservation of water resources. We will get the best model based on the lowest value of Akaike's Information Criterion (AIC) \citep{aic, Akaike}. In this work, we will see that for the Descoberto dam and the Santa Maria dam, the model with the lowest AIC has additive-damped trends and additive seasonality.

In section 2 of this work, we will describe the exponential smoothing models and how to calculate the AIC of the models. Section 3 will present this study's historical series of reservoir levels. In section 4, we will calculate and analyze the results of the AIC criteria of each ES model; we will obtain the appropriate model for each of the historical series, the forecast of these models for the series, and the residual analysis, to confirm if the chosen model is adequate. Finally, section 5 presented the conclusions and perspectives of this work.
.  

\section{Exponential Smoothing}

The data for a time series are $y_{1}$, $y_{2}$, ..., $y_{T}$. A simple method of forecasting this series is to consider the forecast equal to the last observation:
 \begin{equation} 
F_{T+h}=y_{T}.
 \end{equation}
Moreover, we assume that the most recent value is the most important, such that $F_{T+h}$ is the prediction of the variable $y$ for all times above $T$, with $h=1, 2, 3, ...$. The average method assumes that all observations are of equal importance and the data have equal weights when generating predictions: 
\begin{equation} 
F_{T+h}=\frac{1}{T}\sum_{t=1}^{T}y_{T}. 
\end{equation} 

Simple exponential smoothing (SES) is an intermediate forecasting method where the forecast weights decrease exponentially and forecasts for $h>1$ always have the same value, that is:
 \begin{equation} 
F_{T+h}=F_{T+1}=\alpha y_{T}+(1-\alpha)y_{T-1}+\alpha(1-\alpha)^{2}y_{T- 2}+..., 
\end{equation}
$\alpha$ is the smoothing parameter, $0\leq\alpha\leq1$ and the closer to 1 the parameter $\alpha$ is, the greater the weight of recent observations. SES has a greater weight for more recent observations than for past times in time series without trend and seasonality \citep{ses}. Another way to represent the forecast is to define it through the level $l_{T}$, such that: 
\begin{eqnarray} 
F_{T+h}=F_{T+1}=l_{T}\\ 
l_{T}=\alpha y_{T}+(1-\alpha)l_{T-1}. 
\end{eqnarray} 

In 1957 Charles Holt proposed a method for forecasting the time series with the trend. The forecast is through:
 \begin{eqnarray} 
F_{T+h}=l_{T}+hb_{T}\\ 
l_{T}=\alpha y_{T}+(1-\alpha)(l_{T-1}+b_{T-1})\\ 
b_{T}=\beta(l_{T}+b_{T-1})+(1-\beta)b_{T-1}, 
\end{eqnarray} 
with $l_{T}$ the level at time $T$, $b_{T}$ the trend or slope of the series at time $T$, $\alpha$ is the level smoothing parameter, $0\leq\alpha\leq 1$, $\beta$ is the smoothing parameter for the trend and $0\leq\beta\leq1$. The Holt-Winters method (1960) is an extension of the Holt method for series with seasonality; its forecast is calculated through $l_{T}$, $b_{T}$, of the seasonality at the instant $T$, $s_ {T}$, and the smoothing parameters are $\alpha$, $\beta$ and $\gamma$, so: 
\begin{eqnarray} 
F_{T+h}=l_{T}+hb_{T}+s_{T+h-m(k+1)}\\ 
l_{T}=\alpha (y_{T}-s_{T-m})+(1-\alpha)(l_{T-1}+b_{T-1})\\ b_{T}=\beta(l_{T}-l_{T-1})+(1-\beta)b_{T-1}\\
 s_{T}=\gamma(F_{T}-l_{T-1}-b_{T-1})+(1-\gamma)s_{T-m}, 
\end{eqnarray} 
where $m$ is the frequency of seasonality, that is, the number of seasons in the year \citep{book-Hyndman}, and $k$ is the integer part of $(h-1)/m$. This is the additive seasonality version used when the seasonality is constant, and there is the multiplicative seasonality version when the seasonality varies with the level.

In addition to the forecasting methods mentioned above, there is a taxonomy proposed by Pegels \citep{pegels} and employed by Hyndman \citep{hyndman}, which identifies and differentiates forecasting methods according to their trend and seasonality. The Table \ref{tabela1} identifies the fifteen forecasting methods tested in this work.

\begin{table}[!h]
\begin{tabular}{c|c|c|c}
\multirow{1}{*}{Trend} & \multicolumn{3}{c}{Seasonality}       \\
\hline
                           & N-none & A-additive & M-multiplicative \\
 \hline
N-none                     & NN     & NA         & NM               \\
A-additive                 & AN     & AA         & AM               \\
M-multiplicative           & AM     & MA         & MM               \\
AD-additive damped         & ADN    & ADA        & ADM              \\
MD-multiplicative damped   & MDN    & MDA        & MDM             
\end{tabular}
\caption{Taxonomy Exponential Smoothing Methods}\label{tabela1}
\end{table}

The method we choose to get the best model for prediction is by calculating Akaike’s Information Criterion (AIC) and also called the Schwarz criterion:
\begin{equation}
AIC= T\log \left(\frac{SSE}{T}\right)+2c, 
\end{equation}
with $T$ the total number of data used to estimate or train the model, $SSE=\sum_{t=1}^{T}e_{t}^{2}$ the sum of squared errors, $e_{t}=y_{ t}-\hat{y}_{t}$, where $\hat{y}_{t}$ are the model predictions and $c$ the total number of estimated parameters in the model, $c=m\times n_{s}+2\times n_{t}+2+n_{d}$. The variables $n_{s}$, $n_{t}$ and $n_{d}$ can have values of 0 or 1 depending on whether or not the model has seasonality, trend and damping, respectively. For instance, the variable $n_{s}$ is 0 when the model has no seasonality and $n_{s}=1$ when the model has seasonality.

The model chosen will be the one with the smallest AIC since the smaller the AIC, the less information lost from the model. The AIC depends not only on the SSE but also on the total number of estimated parameters. Therefore, models with a higher number of estimated parameters are penalized to the detriment of models with a lower $c$ \citep{book-Hyndman}.

\section{Reservoir water level data}

The reservoirs of the Descoberto, Santa Maria, and Lake Paranoá dams are the main ones responsible for supplying the Federal District. Together, the first two dams provide 83\% of the population \citep{ana}. 

\begin{figure}[!h]
\centering 
\includegraphics[width=10cm]{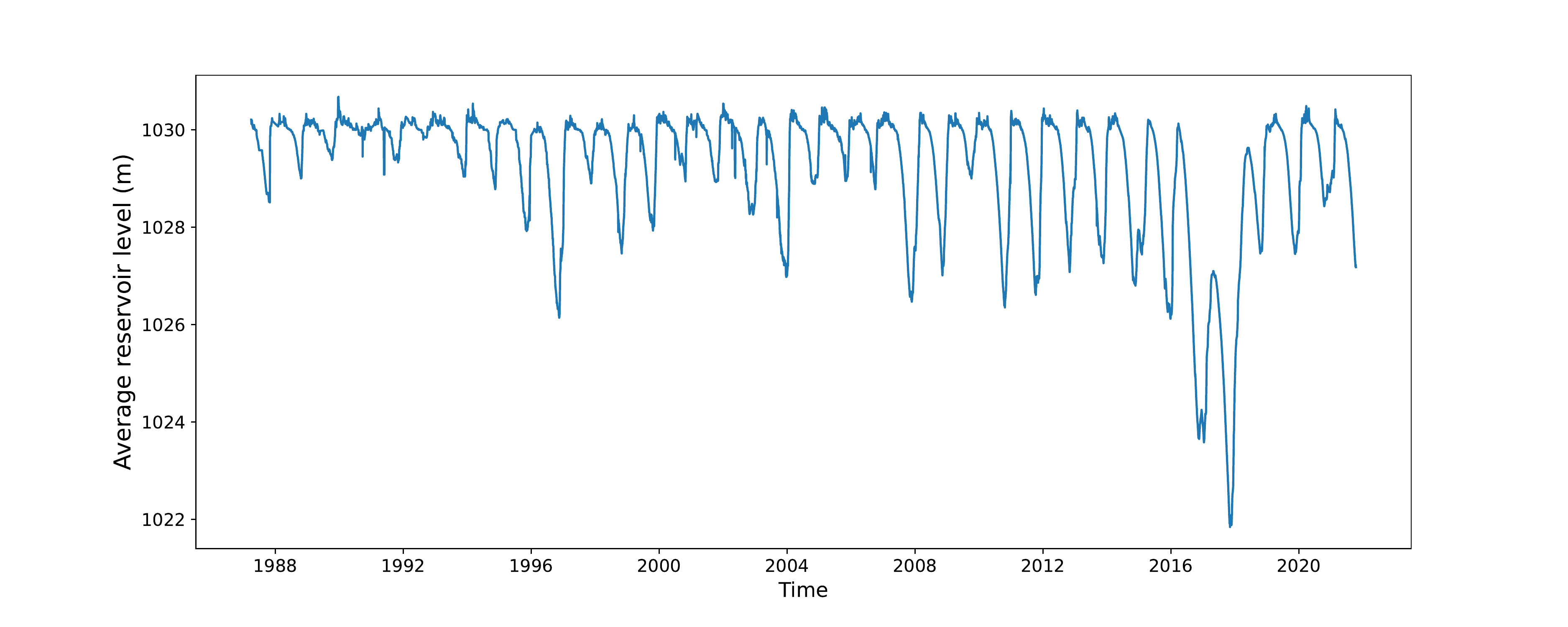} 
\caption{Historical Series of the Descoberto Dam Reservoir (April, 1987- October, 2021)}
\label{descoberto}
\end{figure}

\begin{figure}[!h]
\centering 
\includegraphics[width=10cm]{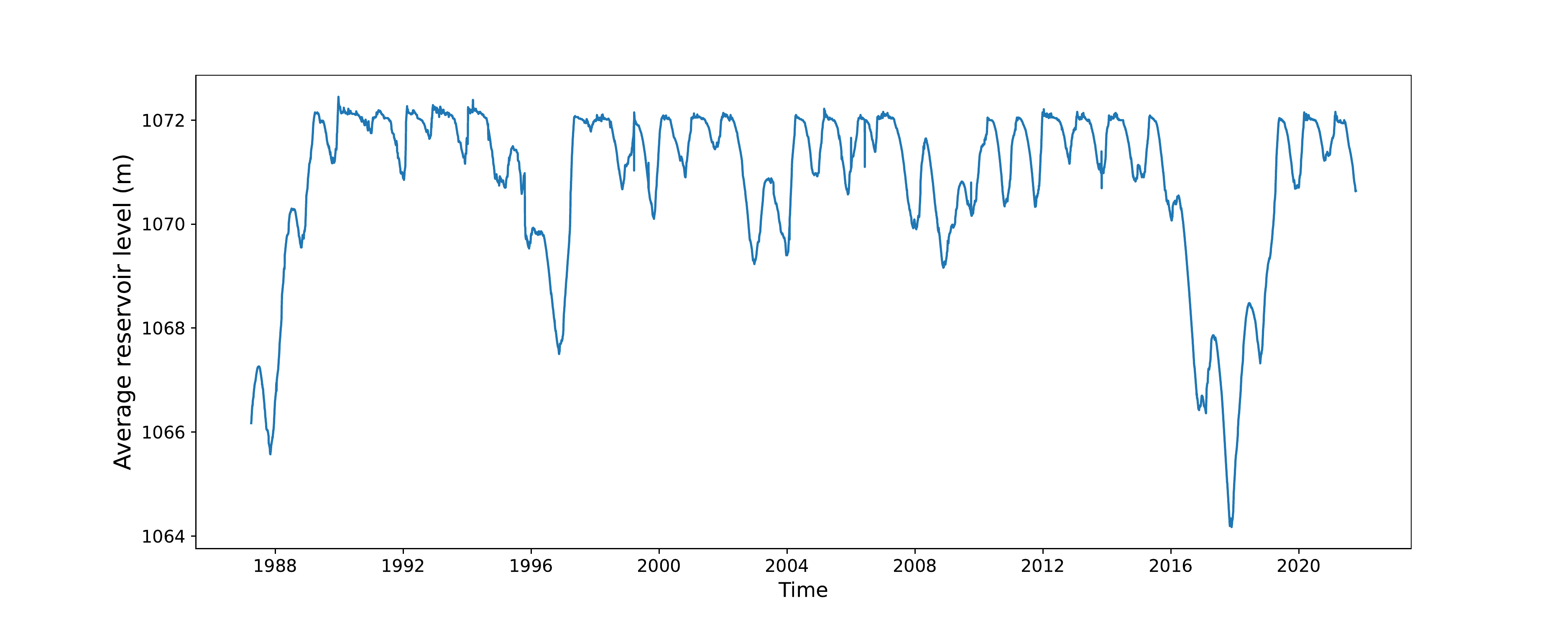}
\caption{Historical Series of the Santa Maria Dam Reservoir (April, 1987- October, 2021)}
\label{santa_maria}
\end{figure}

Obtain the data in \citep{dataset}, have the average daily level of the reservoirs of the Descoberto dam, with 12307 data, of the Santa Maria dam, with 12399 data, and of the Paranoá lake, with 1511 data. However, the small amount of data from Lake Paranoá compared to the other two dams restricts the study to the Descoberto and Santa Maria dams, with historical series from April 1987 to October 2021. Therefore, the historical series of the Descoberto reservoir is shown in Fig. \ref{descoberto}, while the historical series of the Santa Maria reservoir is in Fig. \ref{santa_maria}; in addition, in 2017, there was the lowest average level recorded for the two dams. 

\begin{figure}[!h]
\centering 
\includegraphics[width=12cm]{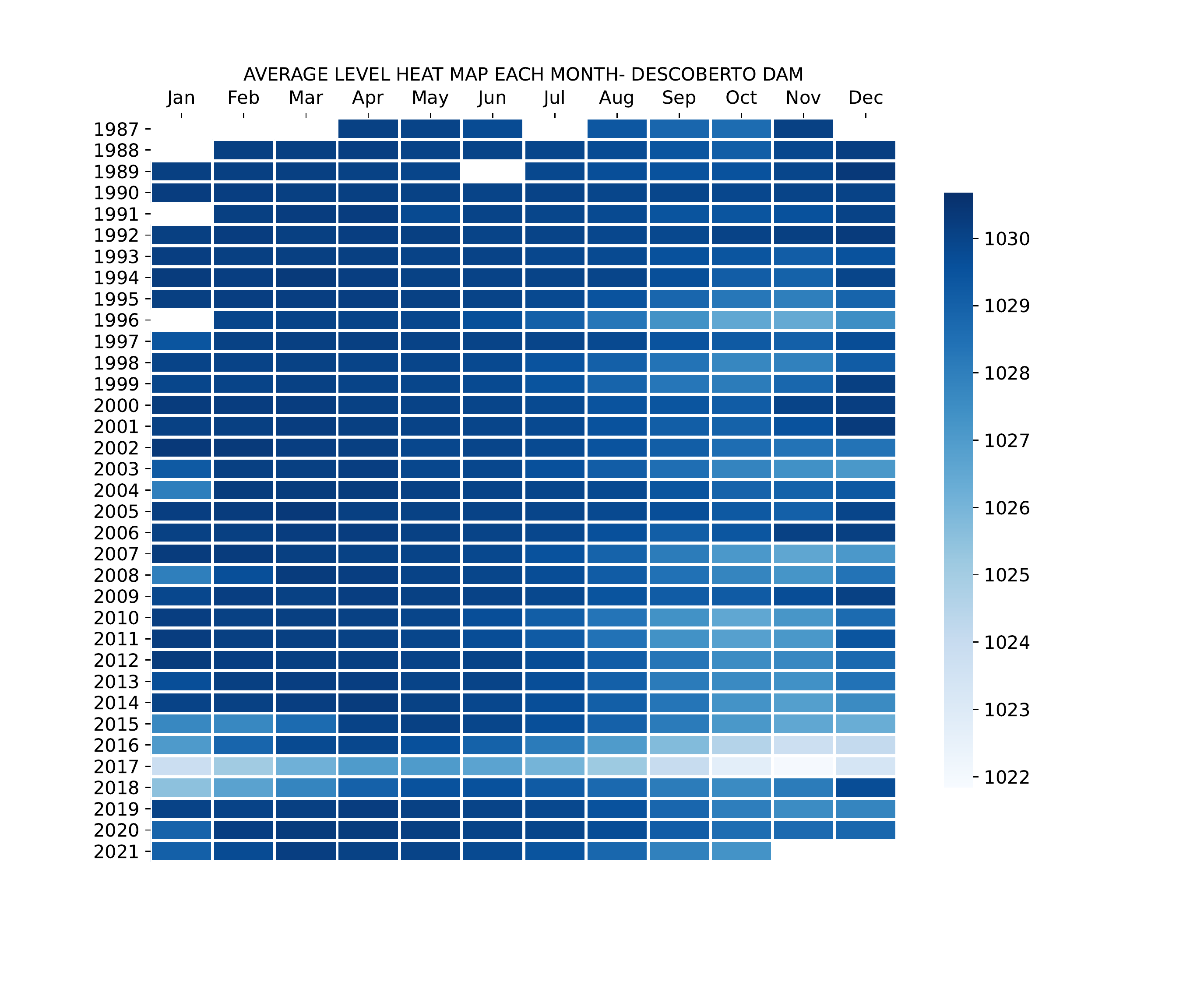} 
\caption{Average level heat map each month- Descoberto Dam (April, 1987- October, 2021)}
\label{descoberto2}
\end{figure}

\begin{figure}[!h]
\centering 
\includegraphics[width=12cm]{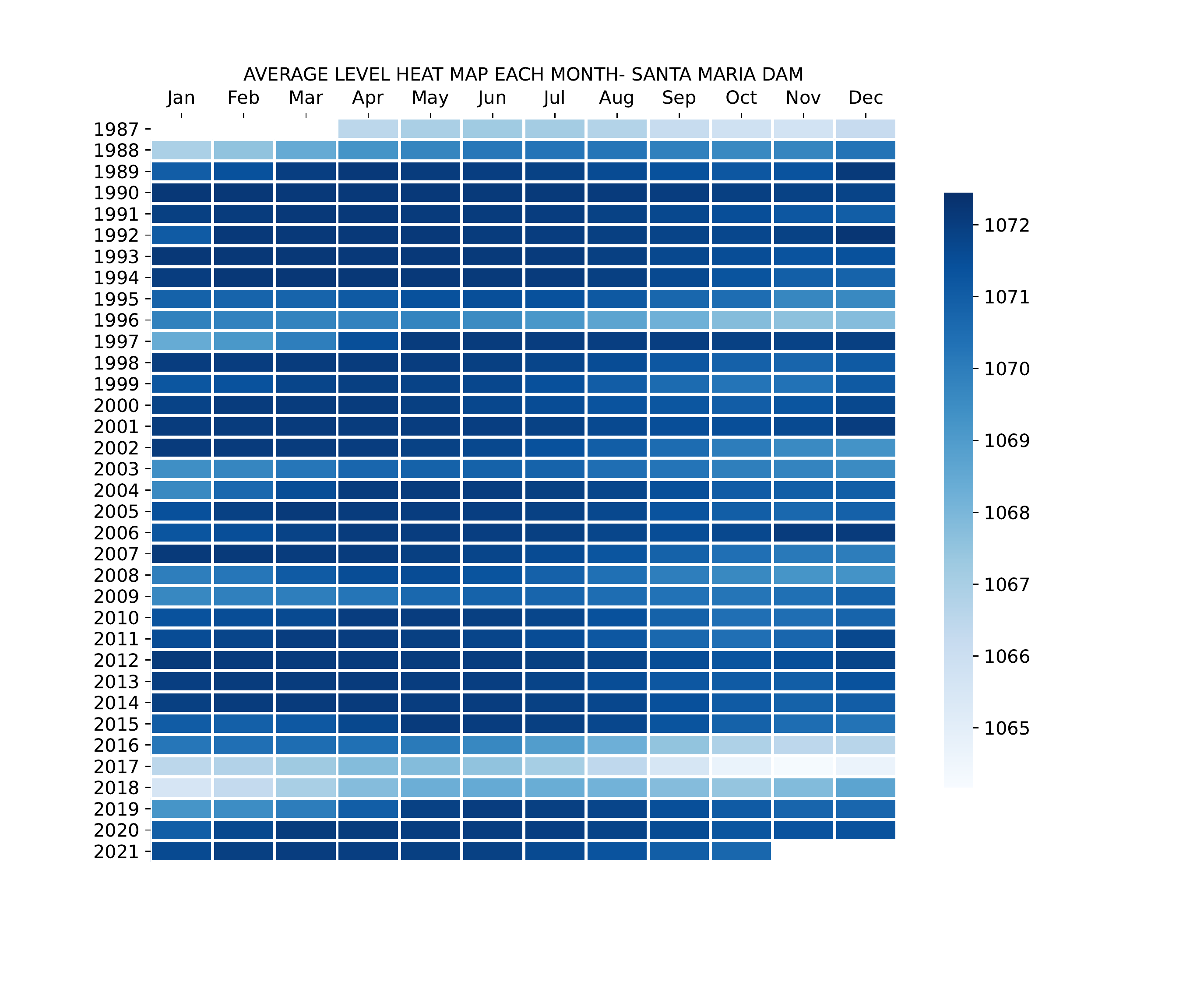}
\caption{Average level heat map each month- Santa Maria Dam (April, 1987- October, 2021)}
\label{santamaria2}
\end{figure}

The Descoberto reservoir reached the minimum level of 1021.84 $m$ on November 07, 2017, and the Santa Maria reservoir reached the minimum level of 1064.17 $m$ on November 25, 2017. As we can see in the heat map of the two dams Fig. \ref{descoberto2} and \ref{santamaria2}, in 2017, there was a dry period, presenting atypical behavior throughout the year.

\section{Results and discussions}

We train the prediction models indicated in Table \ref{tabela1} for the datasets of the Descoberto Dam and the Santa Maria Dam; for both time series, we trained ninety percent of the data, and ten percent were for testing.  Once the AIC is calculated, for each of the models, we obtain Table \ref{tabela2} for the time series of the Descoberto Dam and Table \ref{tabela3} for the Santa Maria Dam.
\begin{table}[!h]
\begin{tabular}{c|ccc}
\hline
Trend                    & \multicolumn{3}{c}{Seasonality}                                                     \\ \hline
                         & \multicolumn{1}{c|}{N-None}     & \multicolumn{1}{c|}{A-Additive} & M-Multiplicative \\ \hline
N-None                   & \multicolumn{1}{c|}{-66702.198} & \multicolumn{1}{c|}{-66915.802} & -65487.895       \\ 
A-Additive               & \multicolumn{1}{c|}{-67801.853} & \multicolumn{1}{c|}{-67370.276} & -65961.216       \\ 
M-Multiplicative         & \multicolumn{1}{c|}{-67801.657} & \multicolumn{1}{c|}{-67369.428} & -65938.873       \\ 
AD-Additive Damped       & \multicolumn{1}{c|}{-67933.668} & \multicolumn{1}{c|}{-67469.059} & -66099.082       \\ 
MD-Multiplicative Damped & \multicolumn{1}{c|}{-67933.873} & \multicolumn{1}{c|}{-7815}      & -66067.551       \\ 
\end{tabular}
\caption{AIC values for the exponential smoothing models applied to the time series of the Descoberto Dam}\label{tabela2}
\end{table}

\begin{table}[!h]
\begin{tabular}{c|ccc}
\hline
Trend                    & \multicolumn{3}{c}{Seasonality}                                                     \\ \hline
                         & \multicolumn{1}{c|}{N-None}     & \multicolumn{1}{c|}{A-Additive} & M-Multiplicative \\ \hline
N-None                   & \multicolumn{1}{c|}{-76368.183} & \multicolumn{1}{c|}{-75995.777} & -72954.198       \\ 
A-Additive               & \multicolumn{1}{c|}{-77546.878} & \multicolumn{1}{c|}{-76029.366} & -73187.861       \\
M-Multiplicative         & \multicolumn{1}{c|}{77726.449}  & \multicolumn{1}{c|}{-55199.861} & -65738.916       \\ 
AD-Additive Damped       & \multicolumn{1}{c|}{-77814.129} & \multicolumn{1}{c|}{-77317.712} & -73054.512       \\ 
MD-Multiplicative Damped & \multicolumn{1}{c|}{-77814.209} & \multicolumn{1}{c|}{136021.169} & -73097.398       \\
\end{tabular}
\caption{AIC values for the exponential smoothing models applied to the time series of the Santa Maria Dam}\label{tabela3}
\end{table}

The time series studied in this work show seasonality; for this reason, we will choose models with seasonality. Therefore, the lowest AIC in ascending order are ADA, AA, MA, and NA for the Descoberto Dam. As for the Santa Maria Dam, the models with the lowest AIC in ascending order are ADA, AA, NA, and AM.
After choosing the four models with the lowest AIC for each time series, we can see in Figs. \ref{descoberto2} and \ref{santamaria2} that 2017 was an atypical period and presented drought. The presence of these data in the model's training can interfere with the choice of the appropriate model, for that we performed a second training of the data before the dry period. In this step, we calculated the AIC in only four exponential smoothing models that performed better in the first test, and we trained eighty-four percent of the data. We obtained the results from Table \ref{tabela4} for the Descoberto Dam and Table \ref{tabela5} for the Santa Maria Dam. We can conclude that the model with the lowest AIC is the ADA model for both series. 
\begin{table}[!h]

\begin{center}
\begin{tabular}{c|c}
\hline
Models & AIC        \\ \hline
ADA    & -62334.936 \\ 
AA     & -62232.353 \\ 
MA     & -62231.139 \\ 
NA     & -61962.861 \\ \hline
\end{tabular}
\end{center}
\caption{AIC values for the ADA, AA, MA and NA models of the Descoberto Dam, training eighty-four percent of the data}\label{tabela4}
\end{table}

\begin{table}[!h]
\begin{center}
\begin{tabular}{c|c}
\hline
Models & AIC        \\ \hline
ADA    & -71364.011 \\ 
AA     & -70546.029 \\ 
NA     & -70482.079 \\
AM     & -68057.242 \\ \hline
\end{tabular}
\end{center}
\caption{AIC values for the ADA, AA, NA and AM models of the Santa Maria Dam, training eighty-four percent of the data}\label{tabela5}
\end{table}
The ADA model is an exponential smoothing model with additive damped trend and additive seasonality, whose recursive formula is:
\begin{eqnarray} 
F_{T+h}=l_{T}+\phi_{h}b_{T}+s_{T+h-m(k+1)}\\ 
l_{T}=\alpha (y_{T}-s_{T-m})+(1-\alpha)(l_{T-1}+\phi b_{T-1})\\
b_{T}=\beta(l_{T}-l_{T-1})+(1-\beta)\phi b_{T-1}\\
s_{T}=\gamma(F_{T}-l_{T-1}-\phi b_{T-1})+(1-\gamma)s_{T-m}\\
\phi_{h}=\phi+\phi^{2}+...+\phi^{h},
\end{eqnarray} 
with $\phi$ the damping parameter, $0<\phi<1$.  

\begin{figure}[!h]
\centering 
\includegraphics[width=10cm]{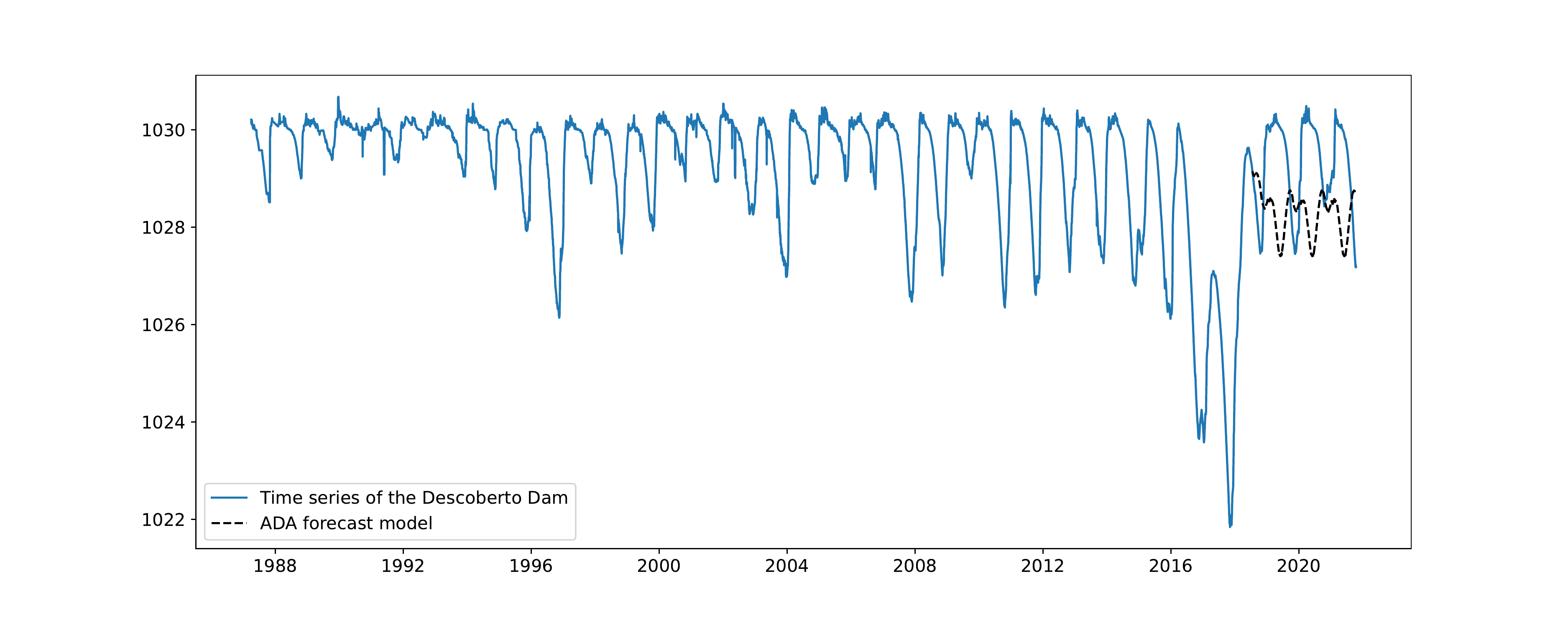} 
\caption{Historical Series of the Descoberto Dam Reservoir (April, 1987- October, 2021) and ADA forecast model.}
\label{descoberto3}
\end{figure}

\begin{figure}[!h]
\centering 
\includegraphics[width=10cm]{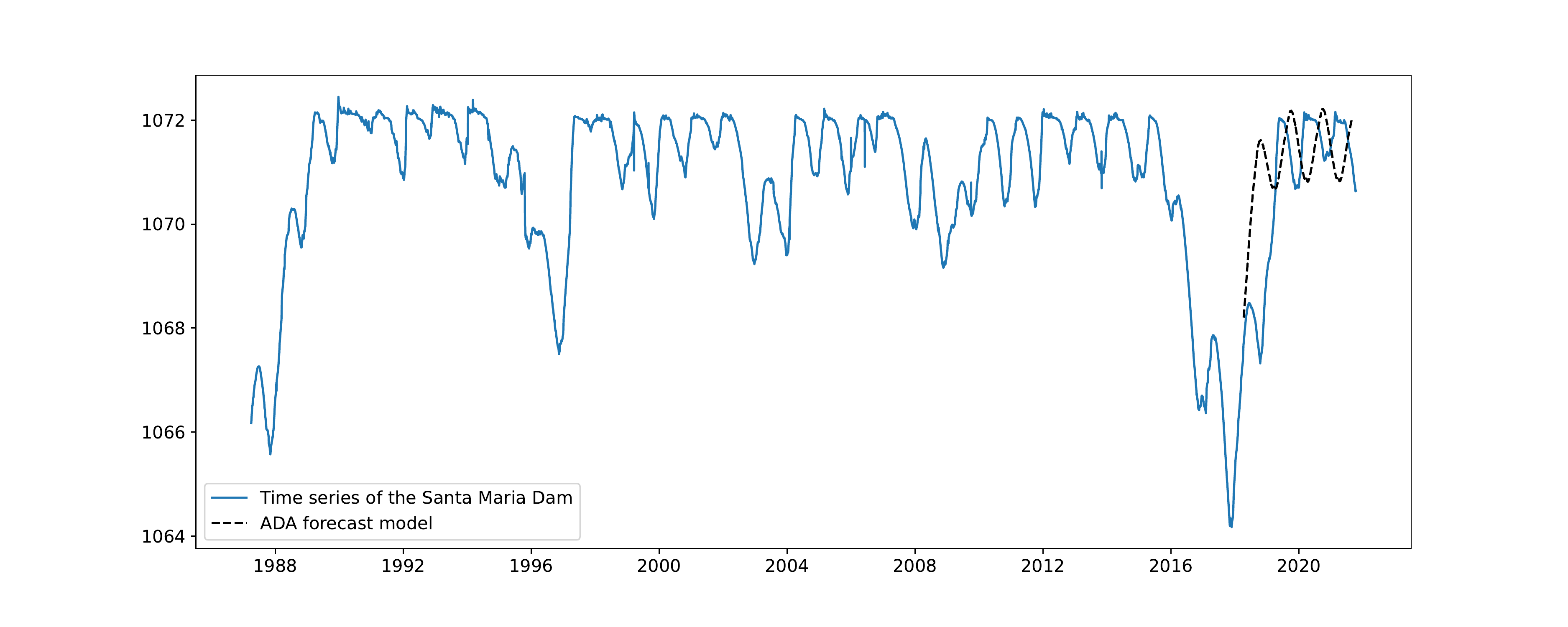} 
\caption{Historical Series of the Santa Maria Dam Reservoir (April, 1987- October, 2021) and ADA forecast model.}
\label{santamaria3}
\end{figure}

Considering the data from the Descoberto Dam and training ninety percent of the data, we obtain that the AIC value of the ADA model is $-67469.059$ and the smoothing parameters are:
\begin{eqnarray}
\alpha=0.931\\
\beta=0.098\\
\gamma=0.002\\
\phi=0.978,
\end{eqnarray}
with $m=365$, $h\in[0,1172]$ e $T=11136$. Figure \ref{descoberto3} shows the historical series of the Descoberto Dam and the ADA forecast model for the ten percent of the data selected and tested.
For the data from the Santa Maria Dam and training ninety percent of the data, we obtain that the AIC value of the ADA model is $-77317.712$ and the smoothing parameters are:
\begin{eqnarray}
\alpha=0.774\\
\beta=0.060\\
\gamma=0.002\\
\phi=0.991,
\end{eqnarray}
with $m=365$, $h\in[0,1240]$ e $T=11159$. Figure \ref{santamaria3} shows the historical series of the Santa Maria Dam and the ADA forecast model for the ten percent of the data selected and tested.

\begin{figure}[!h]
\centering 
\includegraphics[width=10cm]{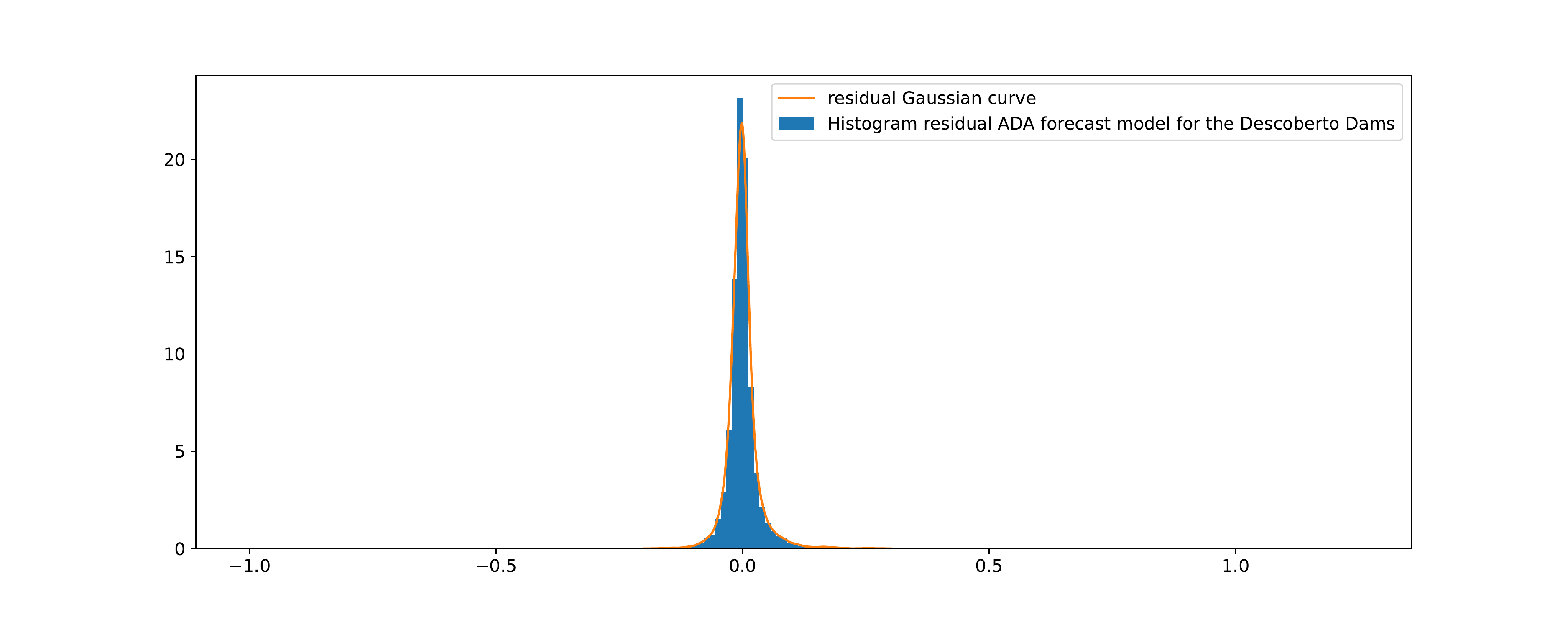}
\includegraphics[width=10cm]{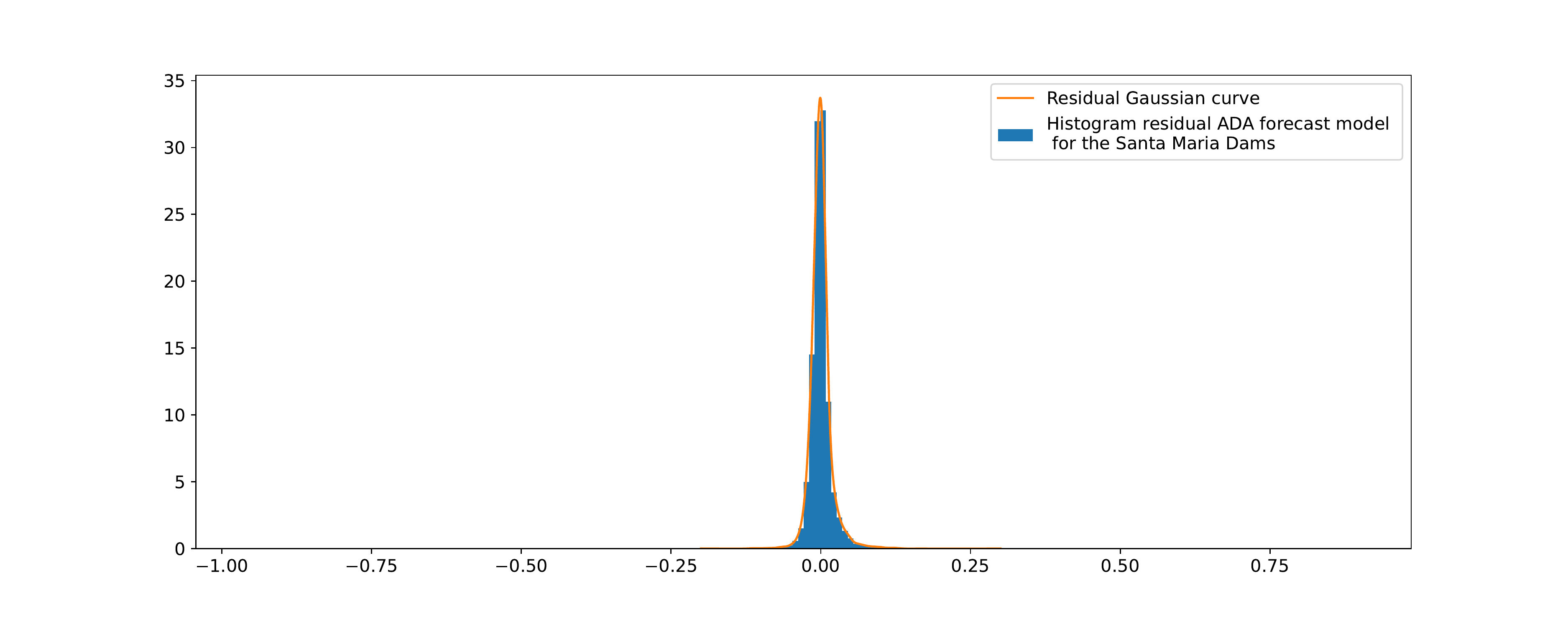} 
\caption{\textit{Above:} Histogram of residues for the Descoberto Dams. \textit{Below:} Histogram of residues for the Santa Maria Dams.}
\label{descoberto4}
\end{figure}

\begin{figure}[!h]
\centering 
\includegraphics[width=10cm]{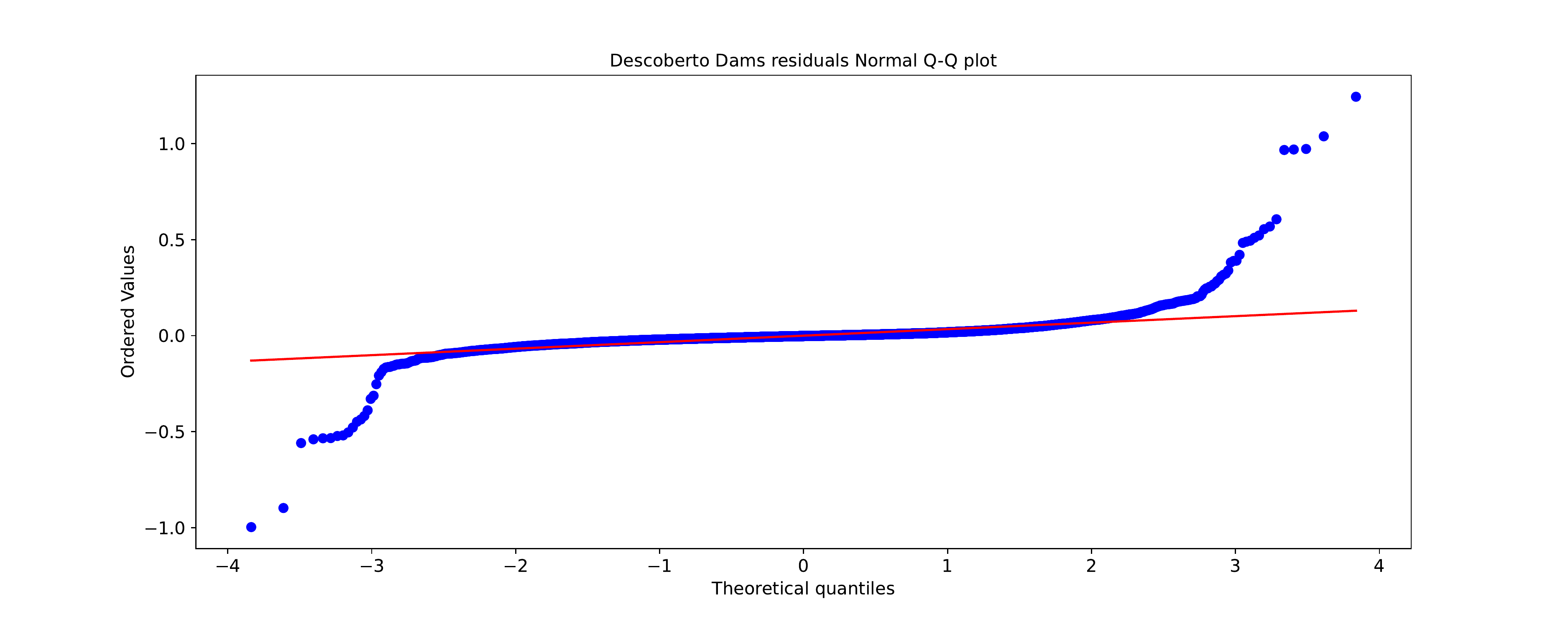}
\includegraphics[width=10cm]{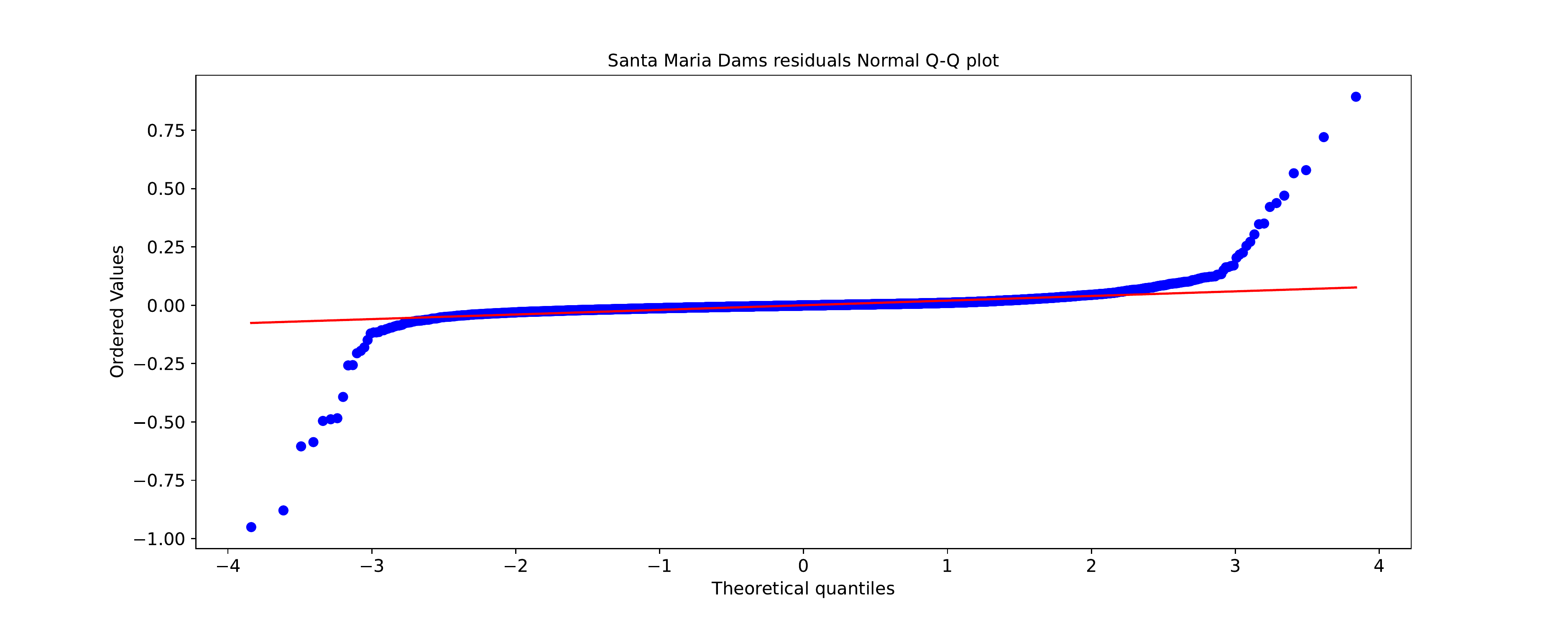} 
\caption{\textit{Above:} Normal Q-Q plot of residues for the Descoberto Dams. \textit{Below:} Normal Q-Q plot of residues for the Santa Maria Dams.}
\label{descoberto5}
\end{figure}

In this work, we use the residual histogram method and the Normal Q-Q plot \citep{q-qplot} to verify that the set of residuals of the model is normality. We draw in Figure \ref{descoberto4} the residue histograms for the Descoberto and Santa Maria dams. The histograms show that the model used for both time series is normalized and corresponds to a Gaussian distribution. Furthermore, in the Normal Q-Q plot method graphs for the two dams in Figure \ref{descoberto5}, we observed the residuals in the graph fall approximately along a straight diagonal line, i.e., the residuals are normality. Therefore, we conclude that the ADA model for the time series of the Descoberto and Santa Maria dams is normal.

\section{Conclusion}

In this work, we study the exponential smoothing models proposed by the Pregels taxonomy, which can present forecasts based on the seasonal component, trend, and level. 

We used the historical series of water levels from the Descoberto and Santa Maria reservoirs, trained for each exponential smoothing model, and calculated the AIC value to obtain the models with the lowest AIC. We performed a new test training the data before the 2017 drought to verify whether the model chosen for the two dams suffered any changes due to the dry period. We confirmed that the model with the lowest AIC in both tests is the ADA model.

Therefore, this study indicates that the ADA model, an exponential smoothing model with damped additive trend and additive seasonality, is the appropriate model to describe the behavior of the two-time series. Based on this model, we also analyzed the residuals to verify whether the model is normal and confirmed this hypothesis.

\pagebreak


\begin{thebibliography}{00}

\bibitem[ANA (2022)]{ana} Agência Nacional de Águas, https://www.ana.gov.br/sar/outros-sistemas-hidricos/df, Accessed December 19, 2022.

\bibitem[dataset (2021)]{dataset} Agência Reguladora de Águas, Energia e Saneamento Básico do Distrito Federal, Monitoramentos dos Níveis de reservatórios, https://www.adasa.df.gov.br/monitoramento/niveis-dos-reservatorios, Accessed October 15, 2021.


\bibitem[Akaike (1974)]{Akaike}Akaike, H. (1974). A new look at the statistical model identification. IEEE Transactions on Automatic Control, 19, 716–723

\bibitem[Brown (1959)]{Brown}Brown, R. G. (1959). Statistical forecasting for inventory control. New York: McGraw Hill

\bibitem[Gardner Jr.(1985)]{Gardner}Gardner Jr., E.S., (1985). Exponential smoothing: The state of the art. Journal of Forecasting, 4, 1-28. 

\bibitem[Gardner Jr. et al.(2008)]{GardnerJr} Gardner Jr., E. S. \& Diaz-Saiz, J. (2008). Exponential smoothing in the telecommunications data. International Journal of Forecasting, 24, 170-174. 

\bibitem[Holt (2004)]{Holt}Holt, C. C. (2004). Forecasting Trends and Seasonal by Exponentially Weighted Averages. International Journal of Forecasting, 20, 5–13.

\bibitem[Hyndman et. al.(2002)]{hyndman}Hyndman, R. J., Koehler, A. B., Snyder, R. D. \& Grose, S.(2002). A state space framework for automatic forecasting using exponential smoothing methods. International Journal of Forecasting, 18, 439–454.
\bibitem[Hyndman, R. J.et al.(2018)]{book-Hyndman} Hyndman, R. J., \& Athanasopoulos, G. (2018). Forecasting: Principles and Practice. (2nd ed.) OTexts.

\bibitem[Kolassa (2011)]{aic}Kolassa, S. (2011). Combining exponential smoothing forecasts using Akaike weights. International Journal of Forecasting, 27, 238–251.

\bibitem[Mahajan \& Tsai (2018)]{sensors}Mahajan, S.,Chen, L-J. \& Tsai,T.-C. (2018).Short-Term PM2.5 Forecasting Using Exponential Smoothing Method: A Comparative Analysis. Sensors, 18(10), 3223.

\bibitem[Ostertagova \& Ostertag (2012)]{ses}Ostertagova, Eva \& Ostertag, Oskar. (2012). Forecasting Using Simple Exponential Smoothing Method. Acta Electrotechnica et Informatica, 12, 62–66, 10.2478/v10198-012-0034-2. 

\bibitem[Pan (2010)]{hw}Pan, Rong. (2010). Holt–Winters Exponential Smoothing. 10.1002/9780470400531.eorms0385. 

\bibitem[Pegels (1969)]{pegels}Pegels, C. C. (1969). Exponential forecasting: some new
variations. Management Science, 12, 311–315.

\bibitem[Raftery (1995)]{bic}Raftery, A. E. (1995). Bayesian model selection in social research.
Sociological Methodology, 25, 111–163.

\bibitem[Siregar et al.(2017)]{Sigerar}Siregar, B., Butar-Butar, I. A., Rahmat, RF., Andayani, U. \& Fahmi, F. (2017). Comparison of Exponential Smoothing Methods in Forecasting Palm Oil Real Production. Journal of Physics: Conference Series, 801, 012004.

\bibitem[Tratara \& Strmčnikb (2017)]{Ferbartrata} Tratara, L. F. \& Strmčnikb, E. (2016). The comparison of Holt-Winters method and Multiple regression method: A case study. Energy, 109, 266-276.

\bibitem[Zafar et al. (2021)]{appli} Zafar, S., Kashif, M., Khan, M. \& Nida, H. (2021).Application of Simple Exponential Smoothing Method for Temperature Forecasting in Two Major Cities Of The Punjab, Pakistan. Agrobiological Records, 4(01),64-68.

\bibitem[Wilk \& Gnanadesikan (1968)]{q-qplot} Wilk, M. B., \& Gnanadesikan, R. (1968). Probability Plotting Methods for the Analysis of Data. Biometrika, 55(1), 1–17. 

\bibitem[Winters (1960)]{Winters}Winters, P. R. (1960). Forecasting sales by exponentially weighted moving averages. Management Science, 6, 324 – 34.


\end{thebibliography}


\end{document}